\documentclass[journal]{IEEEtran}
\usepackage{booktabs}

\usepackage{adjustbox}
\usepackage[colorinlistoftodos]{todonotes}

\def\Gd#1{}
\ifCLASSINFOpdf
\else
\fi
\usepackage{amsmath}
\usepackage{multirow}

\usepackage{url}

\begin{document}

%
\title{Automatic lung segmentation in routine imaging is primarily a data diversity problem, not a methodology problem}
%
%
%

\author{Johannes Hofmanninger, Florian Prayer, Jeanny Pan, Sebastian R\"ohrich, Helmut Prosch, Georg Langs
\thanks{All authors are with the Department of Biomedical Imaging and Image-guided Therapy, Computational Imaging Research Lab, Medical University Vienna. (correspondence: johannes.hofmanninger@meduniwien.ac.at, georg.langs@meduniwien.ac.at)

}}

%
%

\markboth{Accepted for publication in European Radiology Experimental 4, 50 (2020). https://doi.org/10.1186/s41747-020-00173-2}{}
%



\maketitle

\begin{abstract}
Automated segmentation of anatomical structures is a crucial step in image analysis. For lung segmentation in computed tomography, a variety of approaches exist, involving sophisticated pipelines trained and validated on different datasets. However, the clinical applicability of these approaches across diseases remains limited. We compared four generic deep learning approaches trained on various datasets and two readily available lung segmentation algorithms. We performed evaluation on routine imaging data with more than six different disease patterns and three published data sets. Using different deep learning approaches, mean Dice similarity coefficients (DSCs) on test datasets varied not over 0.02. When trained on a diverse routine dataset (n = 36) a standard approach (U-net) yields a higher DSC (0.97 $\pm$ 0.05) compared to training on public datasets such as Lung Tissue Research Consortium (0.94 $\pm$ 0.13, p = 0.024) or Anatomy 3 (0.92 $\pm$ 0.15, p = 0.001). Trained on routine data (n = 231) covering multiple diseases, U-net compared to reference methods yields a DSC of 0.98 $\pm$ 0.03 versus 0.94 $\pm$ 0.12 (p = 0.024).
\end{abstract}


%
\IEEEpeerreviewmaketitle

\section{Background}
The translation of machine learning (ML) approaches developed on specific datasets to the variety of routine clinical data is of increasing importance. As methodology matures across different fields, means to render algorithms robust for the transition from bench to bedside become critical. 

With more than 79 million examinations per year (United States, 2015) \cite{OECD2017}, computed tomography (CT) constitutes an essential imaging procedure for diagnosing, screening and monitoring pulmonary diseases. The detection and accurate segmentation of organs, such as the lung, is a crucial step \cite{mansoor2015}, especially in the context of ML, for discarding confounders outside the relevant organ (e.g., respiration gear, implants or comorbidities) \cite{Zech2018}. 

\begin{figure*}[t]
 \centering
 \includegraphics[width=.99\textwidth]{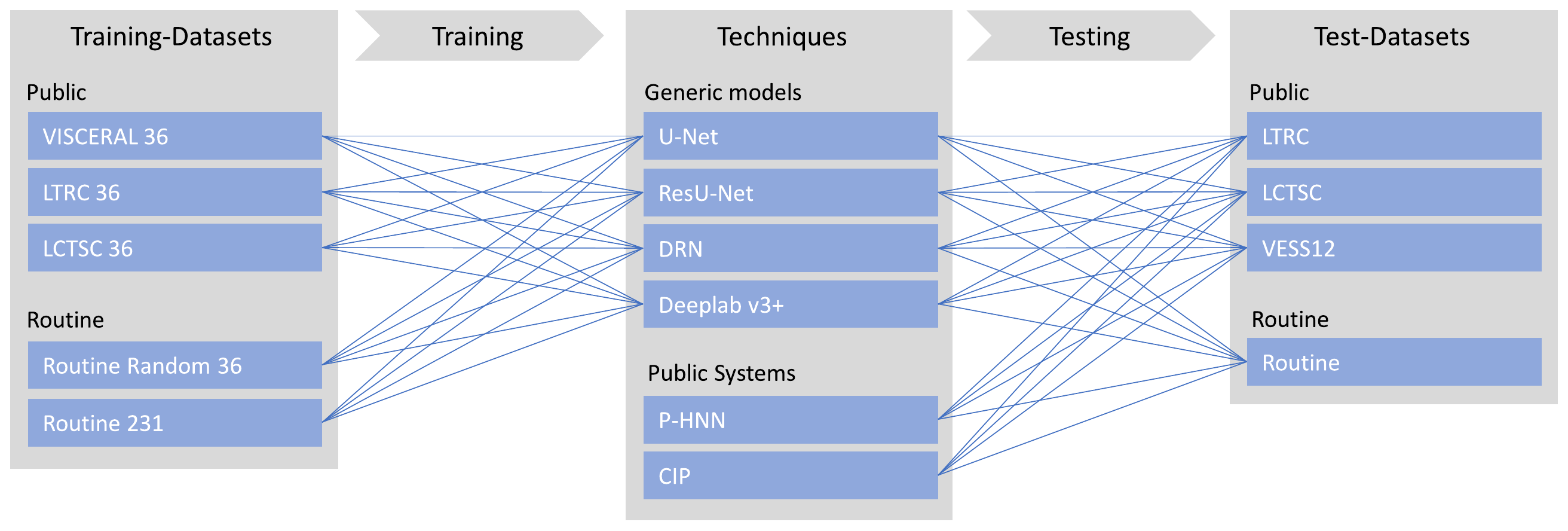}
 \caption{\textbf{Schematic overview of the training and testing performed:} We collected public datasets and two datasets from the routine. We used these datasets to train four generic semantic segmentation models and tested the trained models on public and routine data together with readily available lung segmentation systems}
 \label{fig:figure_1}
\end{figure*}

Automated lung segmentation algorithms are typically developed and tested on limited datasets, covering a limited variability by predominantly containing cases without severe pathology \cite{goksel2015overview} or cases with a single class of disease \cite{yang2018}. Such specific cohort datasets are highly relevant in their respective domain but lead to specialized methods and ML models that struggle to generalize to unseen cohorts when utilized for the task of segmentation. As a consequence, image processing studies, especially when dealing with routine data, still rely on semiautomatic segmentations or human inspection of automated organ masks \cite{Oakden2017,Stein2016}. However, for large-scale data analysis based on thousands of cases, human inspection or any human interaction with single data items, at all, is not feasible. At the same time, disease-specific models are limited with respect to their applicability on undiagnosed cases such as in computer-aided diagnosis or diverse cross-sectional data.
A diverse range of lung segmentation techniques for CT images has been proposed. They can be categorized into rule-based \cite{Korfiatis2007,Hu2001,Chen2011,Pulagam2016}, atlas-based \cite{Sluimer2005,Iglesias2015,li2006}, ML-based \cite{ShanhuiSun2012,agarwala2017,chen2019,softka2011multistage,harrison2017}, and hybrid approaches \cite{Korfiatis2008,Wang2009,Soliman2017,Mansoor2014}. The lung appears as a low-density but high-contrast region on an x-ray-based image, such as CT, so that thresholding and atlas segmentation methods lead to good results in cases with only mild or low density pathologies such as emphysema \cite{Korfiatis2007,Hu2001,Chen2011}. However, disease-associated lung patterns, such as effusion, atelectasis, consolidation, fibrosis, or pneumonia, lead to dense areas in the lung field that impede such approaches. Multi-atlas registration and hybrid techniques aim to deal with these high-density abnormalities by incorporating additional atlases, shape models, and other post-processing steps \cite{Soliman2017,Zhang2001}. However, such complex pipelines are not reproducible without extensive effort if the source code and the underlying set of atlases are not shared. Conversely, trained ML models have the advantage of being easily shared without giving access to the training data. In addition, they are fast at inference time, and scale well when additional training data are available. Harrison et al. \cite{harrison2017} showed that deep-learning-based segmentation outperforms a specialized approach in cases with interstitial lung diseases and provides trained models. However, with some exceptions, trained models for lung segmentation are rarely shared publicly, hampering advances in research. At the same time, ML methods are limited by the training data available, their number, and the quality of the ground-truth annotations.
Benchmark datasets for training and evaluation are paramount to establish comparability between different methods. However, publicly available datasets with manually annotated organs for development and testing of lung segmentation algorithms are scarce. The VISCERAL Anatomy3 dataset \cite{goksel2015overview}, Lung CT Segmentation Challenge 2017 (LCTSC) \cite{yang2018} and the VESsel SEgmentation in the Lung 2012 Challenge (VESSEL12) \cite{rudyanto2012_} provide publicly available lung segmentation data. Yet, these datasets were not published for the purpose of lung segmentation and are strongly biased to either inconspicuous cases or specific diseases neglecting comorbidities and the wide spectrum of physiological and pathological phenotypes. The LObe and Lung Analysis 2011 (LOLA11) challenge published a diverse set of scans for which the ground-truth labels are known only to the challenge organizers \cite{vanRikxoortLObeLOLA11}.
Here, we addressed the following questions: (1) what is the influence of training data diversity on lung segmentation performance; (2) how do inconsistencies in ground truth annotations across data contribute to the bias in automatic segmentation or its evaluation in severely diseased cases; and (3) can a generic deep learning algorithm perform competitively with readily available systems on a wide range of data, once diverse training data are available?

\section{Methods}
We trained four generic semantic segmentation models from scratch on three different public training sets and one training set collected from the clinical routine. We evaluated these models on public test sets and routine data, including cases showing severe pathologies. Furthermore, we performed a comparison of models trained on a diverse routine training set to two published automatic lung segmentation systems, which we did not train, but used as provided. An overview of training and testing performed is given in Fig. \ref{fig:figure_1}.

\subsection{Routine data extraction}
The local ethics committee of the Medical University of Vienna approved the retrospective analysis of the imaging data. We collected representative training and evaluation datasets from the picture archiving and communication system of a university hospital radiology department. We included inpatients and outpatients who underwent a chest CT examination during a period of 2.5 years, with no restriction on age, sex, or indication. However, we applied minimal inclusion criteria with regard to imaging parameters, such as primary and original DICOM tag, number of slices in a series $\geq 100$, sharp convolution kernel, and series description included one of the terms lung, chest, or thorax. If multiple series of a study fulfilled these criteria, the one series with the highest number of slices was used assuming lower inter-slice distance or larger field of view. Scans which did not or only partially showed the lung or scans with patients in lateral position were disregarded. In total, we collected more than 5,300 patients (examined during the 2.5-year period), each represented by a single CT series.

\subsection{Training datasets}
To study training data diversity, we assembled four datasets with an equal number of patients (n = 36) and slices (n = 3,393). These individual datasets were randomly extracted from the public VISCERAL Anatomy3 (VISC-36), LTRC (LTRC-36), and LCTSC (LCTSC-36) datasets, and from the clinical routine (R-36). 
In addition, we carefully selected a large representative training dataset from the clinical routine using three sampling strategies: (1) random sampling of cases (n = 57); (2) sampling from image phenotypes \cite{hofmanninger2016} (n = 71) (the exact methodology for phenotype identification was not in the scope of this work); and (3) manual selection of edge cases with severe pathologies, such as fibrosis (n = 28), trauma (n = 20), and other cases showing extensive ground-glass opacity, consolidations, fibrotic patterns, tumours, and effusions (n = 55). In total, we selected 231 cases from routine data for training (hereafter referred to as R-231). Besides biology, technical acquisition parameters are an additional source of appearance variability. The R-231 dataset contains scans acquired with 22 different combinations of scanner manufacturer, convolution kernel, and slice thickness. While the dataset collected from the clinical routine showed a high variability in lung appearance, cases that depict the head or the abdominal area are scarce. To mitigate this bias toward slices that showed the lung, we augmented the number of non-lung slices in R-231 by including all slices which did not show the lung from the Anatomy3 dataset. Table \ref{tbl:trainingds} lists the training data collected.

\begin{table*}[!t]
\centering
\begin{tabular}{lllll}
\toprule
Abbreviation &               Name & Number of volumes & Number of slices-L & Total number of slices\\
\midrule
R-36 &     Routine Random &    36 &    3393 &    3393\\
VISC-36 &           VISCERAL &    36 &    3393 &    3393\\
LTRC-36 & LTRC & 36 & 3393 & 3393\\
LCTSC-36 &              LCTSC &    36 &    3393 &    3393\\
R-213 &  Routine 231 Cases &   231 &   62224 &    108248\\
\bottomrule
\hspace{.05em}
\end{tabular}
\caption{\textbf{Datasets used to train semantic segmentation models: }The number of volumes, the number of slices that showed the lung (slices-L), and the total number of slices are listed. Visceral, LTRC, and LCTSC are public datasets; R-36 and R-231 are images from the routine database of a radiology department}
\label{tbl:trainingds}
\end{table*}

\begin{figure*}[t!]
 \centering
 \includegraphics[width=.98\textwidth]{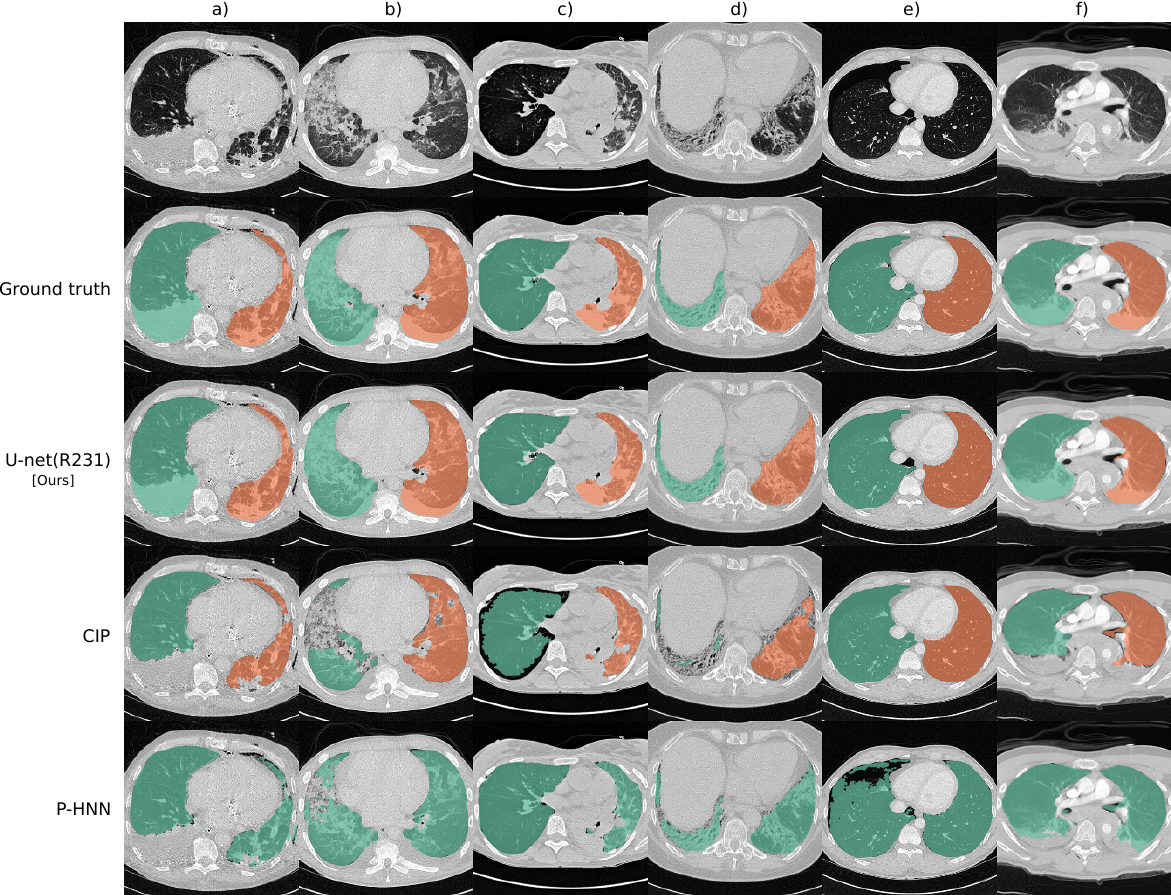}
 \caption{\textbf{Segmentation results for selected cases from routine data:}  Each column shows a different case. Row 1 shows a slice without lung masks, row 2 shows the ground truth, and rows 3 to 5 show automatically generated lung masks. Effusion, chest tube, and consolidations \textbf{(a)}; small effusions, ground-glass and consolidation \textbf{(b)}; over-inflated (right) and poorly ventilated (left), atelectasis \textbf{(c)}; irregular reticulation and traction bronchiectasis, fibrosis \textbf{(d)}; pneumothorax \textbf{(e)}; and effusions and compression atelectasis (trauma) \textbf{(f)}}
 \label{fig:qual_routine}
\end{figure*}

\subsection{Test datasets}
For testing, we randomly sampled 20 cases from the routine database that were not part of the training set and 15 cases with specific anomalies: atelectasis (n = 2), emphysema (n = 2), fibrosis (n = 4), mass (n = 2), pneumothorax (n = 2) and trauma (n = 3)]. In addition, we tested on cases from the public LTRC, LCTSC and VESSEL12 datasets, which were not used for training. Table \ref{tbl:testds} lists the test data collected. Further, we calculated results on a combined dataset composed of the individual test sets (All(L), n = 191). In addition, we report all test cases combined without the LTRC and LCTSC data considered (All, n = 62). The rationale behind this is that the LTRC test dataset contains 105 volumes and dominates the average scores, and the LCTSC dataset contains multiple cases with tumours and effusions that are not included in the ground-truth masks. Thus, an automated segmentation that includes these areas yields a lower score, distorting and misrepresenting the combined results.

\begin{table*}[!t]
\centering
\begin{tabular}{lllll}
\toprule
    Abbreviation &                 Description & Number of volumes  & Number of slices-L & Total number of slices\\
\midrule
      RRT &  Routine Random Test &    20 & 5788 & 7969\\
     LTRC & LTRC & 105 & 44784 & 51211\\
    LCTSC &                LCTSC &    24  & 2063 & 3675\\
   VESS12 &             VESSEL12 &    20  & 7251 & 8593\\
     Ath. &         Atelecatasis &     2  & 395 & 534\\
    Emph. &            Emphysema &     2  & 688 & 765\\
    Fibr. &             Severe fibrosis &     4  & 1192 & 1470\\
    Mass* &                 Mass &     2  & 220 & 273\\
     PnTh &        Pneumo Thorax &     2  & 814 & 937\\
   Trauma &               Trauma/Effusions &     3  & 911 & 2225\\
 Normal** &               Normal (Large field of view) &     7 & 1180 & 5301\\
\bottomrule
 & Total & 191 & 65286 & 82953 \\
\end{tabular}
\caption{\textbf{Test datasets used to evaluate the performance of lung segmentation algorithms: }The number of volumes, the number of slices that showed the lung (slices-L), and the total number of slices are listed. LTRC, LCTSC, and VESS12 are cases from the respective public dataset that were not used for training *Two cases from the publicly available Lung1 dataset **Four cases from the publicly available Visceral Anatomy 3 dataset}
\label{tbl:testds}
\end{table*}

\subsection{Ground truth annotations}
Ground-truth labelling on the routine data was bootstrapped by training of a lung segmentation algorithm (U-net) on the Anatomy3 dataset. The preliminary masks were iteratively corrected by four readers: two radiologists with 4 and 5 years of experience in chest CT and two medical image analysis experts with 6 and 2 years of experience in processing chest CT scans. The model for the intermediate masks was iteratively retrained after 20–30 new manual corrections were performed using the ITK-Snap software \cite{itksnap}.

\subsection{Segmentation methods}
We refrained from developing specialized methodology but utilized generic state-of-the-art deep-learning, semantic segmentation architectures that were not specifically proposed for lung segmentation. We trained these “vanilla” models without modifications and without pre-training on other data. We considered the following four generic semantic segmentation models: U-net; ResU-net; Dilated Residual Network-D-22; and Deeplab v3+. \\\\
\textbf{U-net -} Ronneberger et al. \cite{Ronneberger2015} proposed the U-net for the segmentation of anatomic structures in microscopy images. Since then, it has been used for a wide range of segmentation tasks and various modified versions have been studied \cite{Zhou2017,isensee2019}. We utilized the U-net with the only adaption being batch-normalization \cite{batchnorm2015} after each layer. \\
\textbf{ResU-net -} Residual connections have been proposed to facilitate the learning of deeper networks \cite{schmid2015,He2016}. The ResU-net model includes residual connections at every down- and up-sampling block as a second adaptation to the U-net, in addition to batch-normalization. \\
\textbf{Dilated Residual Network-D-22 -} Yu et al. \cite{yu2015} proposed dilated convolutions for semantic image segmentation and adapted deep residual networks \cite{He2016} with dilated convolutions to perform semantic segmentations on natural images. Here, we utilized the Dilated Residual Network-D-22 model, as proposed by Yu et al. \cite{yu2017}. \\
\textbf{Deeplab v3+ -} Deeplab v3 combines dilated convolutions, multi-scale image representations, and fully-connected conditional random fields as a post-processing step. Deeplab v3+ includes an additional decoder module to refine the segmentation. Here, we utilized the Deeplab v3+ model as proposed by Chen et al. \cite{Chen2018}. \\\\
We compared the trained models to two readily available reference methods: the Progressive Holistically Nested Networks (P-HNN) and the Chest Imaging Platform (CIP). The P-HNN has been proposed by Harrison et al. \cite{harrison2017} for lung segmentation. The upon request available model was trained on cases from the public LTRC dataset (618 cases) and other cases with interstitial lung diseases or infectious diseases (125 cases). The CIP provides an open source lung segmentation tool based on thresholding and morphological operations \cite{ChestCIP_}.

\subsection{Experiments}
We determined the influence of training data variability (especially public datasets versus routine) on the generalizability to other public test-datasets, and, specifically, to cases with a variety of pathologies. To establish comparability, we limited the number of volumes and slices to match the smallest dataset from LCTSC, with 36 volumes and 3,393 slices. During this experiment, we considered only slices that showed the lung (during training and testing) to prevent a bias induced by the field of view. For example, images in VISCERAL Anatomy 3 showed either the whole body or the trunk, including the abdomen, while other datasets, such as LTRC, LCTSC, or VESSEL12 contained only images limited to the chest.
Further, we compared the generic models trained the R-231 dataset to the public available systems CIP and P-HNN. For this comparison, we processed the full volumes. The CIP algorithm was shown to be sensitive to image noise. Thus, if the CIP algorithm failed, we pre-processed the volumes with a Gaussian filter kernel. If the algorithm still failed, the case was excluded for comparison. The trained P-HNN model does not distinguish between left and right lung. Thus, evaluation metrics were computed on the full lung for masks created by P-HNN. In addition to evaluation on publicly available datasets and methods, we performed an independent evaluation of our lung segmentation model by submitting solutions to the LOLA11 challenge for which 55 CT scans are published but ground-truth masks are available only to the challenge organizers. Prior research and earlier submissions suggest inconsistencies in the ground truth of the LOLA11 dataset, especially with respect to pleural effusions \cite{Mansoor2014}. We specifically included effusions in our training datasets. To account for this discrepancy and improve comparability we submitted two solutions: first, masks as yielded by our model and alternatively, with subsequently removed dense areas from the lung masks. The automatic exclusion of dense areas was performed by simple thresholding of values between -50 $<$ HU $<$ 70 and morphological operations.
Studies on lung segmentation usually use overlap- and surface-metrics to assess the automatically generated lung mask against the ground truth. However, segmentation metrics on the full lung can only marginally quantify the capability of a method to cover pathological areas in the lung as pathologies may be relatively small compared to the lung volume. Carcinomas are an example of high-density areas that are at risk of being excluded by threshold- or registration-based methods when they are close to the lung border. We utilized the publicly available, previously published Lung1 dataset \cite{Aerts2014} to quantify the model’s ability to cover tumour areas within the lung. The collection contains scans of 318 non-small-cell lung cancer patients before treatment, with a manual delineation of the tumours. In this experiment, we evaluated the overlap proportion of tumour volume covered by the lung mask.

\subsection{Implementation details}
We aimed to achieve a maximum of flexibility with respect to the field of view (from partially visible organ to whole-body) and to enable lung segmentation without prior localization of the organ. To this end, we performed segmentation on the slice level. That is, for volumetric scans, each slice was processed individually. We segmented the left and right lung (individually labelled), excluded the trachea and specifically included high density anomalies such as tumour and plural effusions. During training and inference, the images were cropped to the body region using thresholding and morphological operations and rescaled to a resolution of 256 $\times$ 256 pixels. Prior to processing, Hounsfield units were mapped to the intensity window [-1024; 600] and normalized to the 0-1 range. During training, the images were augmented by random rotation, non-linear deformation and Gaussian noise. We used stratified mini-batches of size 14 holding 7 slices showing the lung and 7 slices which don’t show the lung. For optimization, we used Stochastic Gradient Decent with momentum.

\subsection{Statistical methods}
Automatic segmentations were compared to the ground truth for all test datasets using the following evaluation metrics, as implemented by the Deepmind surfacedistance python module \cite{2018LibraryTasks}. While segmentation was performed on two-dimensional slices, evaluation was performed on the three-dimensional volumes. If not reported differently, the metrics were calculated for the right and left lung separately and then averaged. For comparison between results, paired t-tests have been performed.

\subsubsection{Dice coefficient (DSC)}
The Dice coefficient or Dice score is a measure of overlap:
\begin{equation}
    D(X,Y)=\frac {2|X \cap Y|}{|X|+|Y|}
\end{equation}
where $X$ and $Y$ are two alternative labelings, such as predicted and ground-truth lung masks.

\subsubsection{Robust Hausdorff distance (HD95)}
The directed Hausdorff distance is the maximum distance over all distances from points in surface $X_s$ to their closest point in surface $Y_s$. In mathematical terms, the directed robust Hausdorff distance is given as:
\begin{equation}
    \overrightarrow{H}({X_s,Y_s}) = \underset{x\in X_s}{P_{95}}\left(\min_{y\in{Y_s}}d(x,y)\right)   
\end{equation}
where $P_{95}$ denotes the $95th$ percentile of the distances. Here, we used the symmetric adaptation:
\begin{equation}
   H({X_s,Y_s}) = \max\left(\overrightarrow{H}({X_s,Y_s}),\overrightarrow{H}({Y_s,X_s})\right)
\end{equation}

\subsubsection{Mean surface distance (MSD)}
The MSD is the average distance of all points in surface $X_s$ to their closest corresponding point in surface $Y_s$:

\begin{equation}
    \overrightarrow{MSD}(X_s,Y_s) = \frac {1}{|X|} \sum_{x \in X_s}\min_{y \in Y_s}d(x,y)
\end{equation}
Here, we used the symmetric adaptation:
\begin{equation}
    {MSD}(X_s,Y_s) = \max\left(\overrightarrow{MSD}(X_s,Y_s),\overrightarrow{MSD}(Y_s,X_s)\right)
\end{equation}

\begin{figure*}[t!]
 \centering
 \includegraphics[width=.99\textwidth]{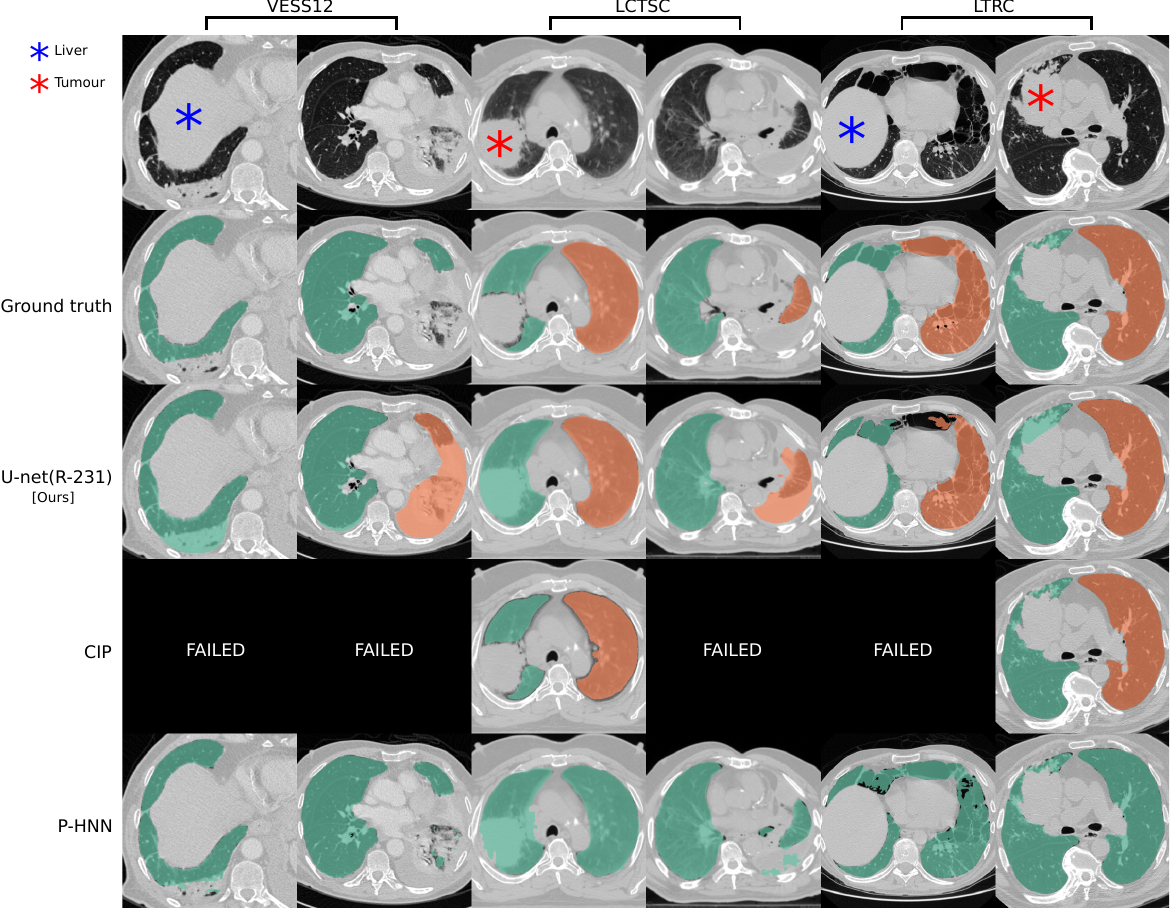}
 \caption{\textbf{Ground truth annotations in public datasets lack coverage of pathologic areas:}  Segmentation results for cases in public datasets where the masks generated by our U-net(R-231) yielded low Dice similarity coefficients when compared to the ground truth. Note that public datasets often do not include high-density areas in the segmentations. Tumours in the lung area should be included in the segmentation while the liver should not.}
 \label{fig:qual_pub}
\end{figure*}

\section{Results}
\begin{figure}[t]
 \centering
 \includegraphics[width=.49\textwidth]{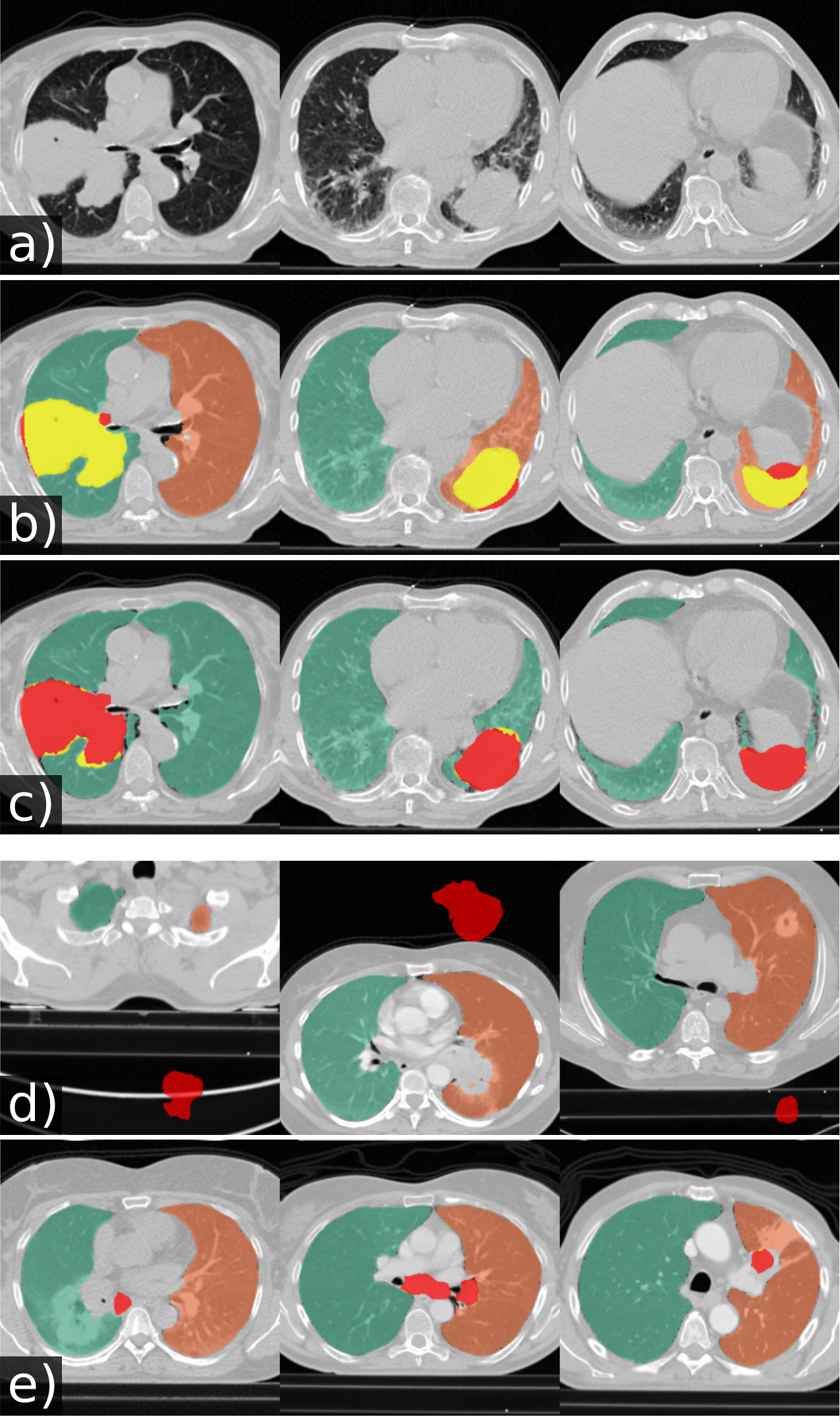}
 \caption{\textbf{U-net trained on routine data covered more tumor area compared to reference methods.} Box- and swarm-plots showing the percentage of tumor volume (318 cases) covered by the lung masks generated by different methods.}
 \label{fig:tumor_box}
\end{figure}

\begin{figure}[t!]
 \centering
 \includegraphics[width=.49\textwidth]{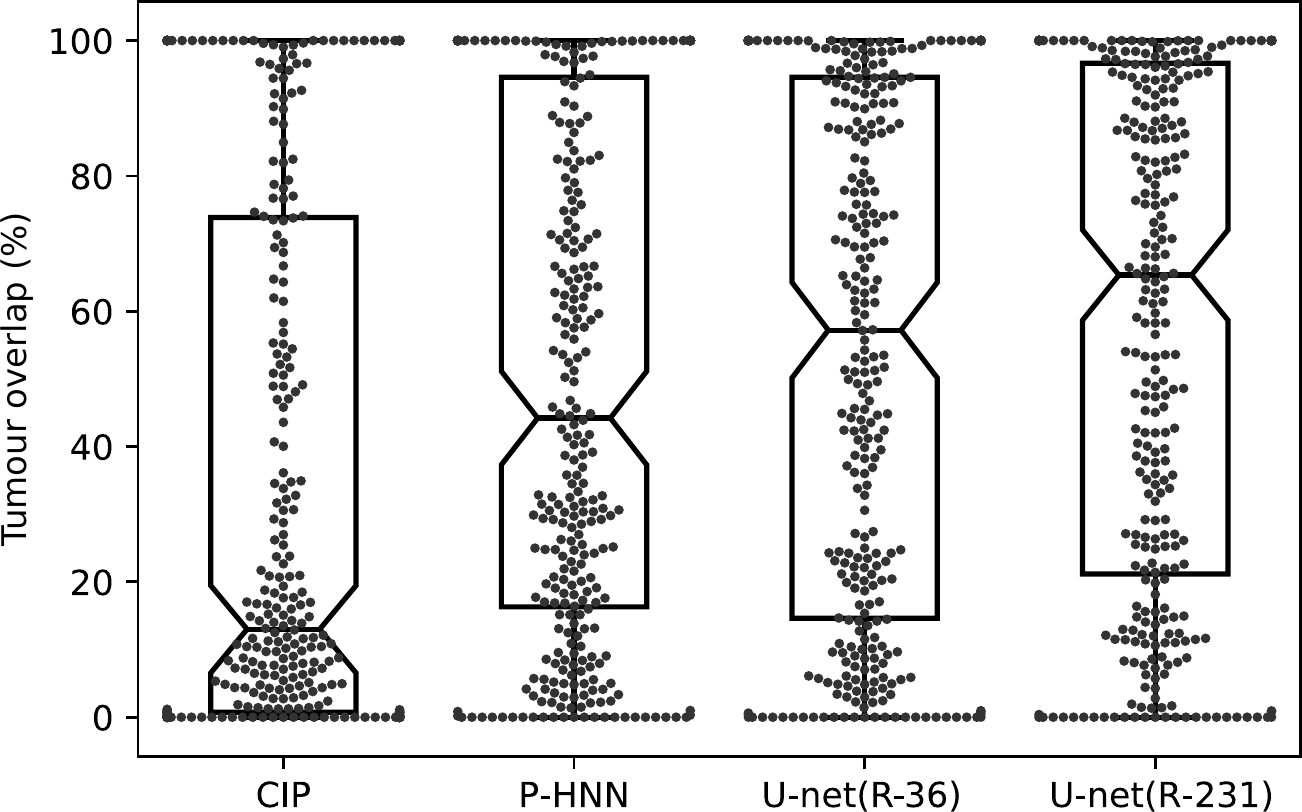}
 \caption{\textbf{Qualitative results of automatically generated lung masks for tumor cases.} Yellow: tumor area covered by the lung mask. Red: tumor area not covered by the lung mask. \textbf{(a)} Lung masks generated by our U-net(R-213). \textbf{(b)} Lung masks generated by P-HNN. \textbf{(c)} Corrupted tumor segmentations in the Lung1 dataset. \textbf{(d)} Cases with poor tumor overlap of lung masks generated by U-net(R-213).}  
 \label{fig:show_tumor}
\end{figure}

Models trained on routine data achieve improved evaluation scores compared to models trained on publicly available study data. U-net, ResU-net, and Deeplab v3+ models, when trained on routine data (R-36), yielded the best evaluation scores on the merged test dataset (All, n = 62). The U-net yields mean DSC, HD95, and MSD scores of 0.96 $\pm$ 0.08, 9.19 $\pm$ 18.15, 1.43 $\pm$ 2.26 when trained on R-36 [U-net(R-36)] and 0.92 $\pm$ 0.14, 13.04 $\pm$ 19.04, 2.05 $\pm$ 3.08 when trained on VISC-36 (R-36 versus VISC-36, p = 0.001, 0.046, 0.007) or 0.94 $\pm$ 0.13, 11.09 $\pm$ 22.9, 2.24 $\pm$ 5.99 when trained on LTRC-36 (R-36 versus LTRC-36, p = 0.024, 0.174, 0.112). This advantage of routine data for training is also reflected in results using other combinations of model architecture and training data. Table \ref{tbl:results_lonly} lists the evaluation results in detail. 
We determined that the influence of model architecture is marginal compared to the influence of training data. Specifically, the mean DSC does not vary for more than 0.02 when the same combination of training and test set was used for different architectures (Table \ref{tbl:results_lonly}).
Compared to readily available trained P-HNN model, the U-net trained on the R-231 routine dataset [U-net(R-231)] yielded mean DSC, HD95, and MSD scores of 0.98 $\pm$ 0.03, 3.14 $\pm$ 7.4, 0.62 $\pm$ 0.93 versus 0.94 $\pm$ 0.12, 16.8 $\pm$ 36.57, 2.59 $\pm$ 5.96 (p = 0.024, 0.004, 0.011) merged test dataset (All, n = 62). For comparison with the CIP-algorithm, only volumes for which the algorithm did not fail were considered. On the merged dataset (All, N=62) the algorithms yielded mean DSC, HD95, and MSD scores of 0.98 $\pm$ 0.01 ,1.44 $\pm$ 1.09 ,0.35 $\pm$ 0.19 for the U-net(R213) compared to 0.96 $\pm$ 0.05, 4.65 $\pm$ 6.45, 0.91 $\pm$ 1.09 for CIP (p = 0.001, $<$0.001, $<$0.001). Detailed results are given in Table \ref{tbl:comp_models}. Fig. \ref{fig:qual_routine} shows qualitative results for cases from the routine test sets and Fig. \ref{fig:qual_pub} shows cases for which the masks generated by the U-net(R-231) model yielded low DSCs when compared to the ground truth.
We created segmentations for the 55 cases of the LOLA11 challenge with the U-net(R-231) model. The unaltered masks yielded a mean overlap score of 0.968 and with dense areas removed 0.977. 
Table \ref{tbl:tumoroverlapp} and Fig. \ref{fig:tumor_box} show results for tumour overlap on the 318 volumes of the Lung1 dataset. U-net(R-231) covered more tumour volume mean/median compared to P-HNN (60\%/69\% versus 50\%/44\%, p $<$ 0.001) and CIP (34\%/13\%). Qualitative results for tumour cases for U-net(R-231) and P-HNN are show in Figs. \ref{fig:show_tumor}b, c. We found that 23 cases of the Lung1 dataset had corrupted ground-truth annotation of the tumours (Fig. \ref{fig:show_tumor}d). Fig. \ref{fig:show_tumor}e shows cases with little or no tumour overlap achieved by U-net(R-231).

\begin{table*}[t]
\begin{adjustbox}{width=.99\textwidth,center}
\begin{tabular}{llrrrrrrrrrrrllll}
{} & {} &   \multicolumn{11}{c}{Test datasets (DSC) for lung slices only} &  \multicolumn{2}{c}{DSC$\pm$SD}  &    \multicolumn{1}{c}{HD95(mm)$\pm$SD} &   \multicolumn{1}{c}{MSD(mm)$\pm$SD} \\
\cmidrule(r{0.7em}){3-13} \cmidrule(r{0.7em}l{0.7em}){14-15} \cmidrule(r{0.7em}l{0.7em}){16-16} \cmidrule(r{0.7em}l{0.7em}){17-17}
{} & {} &   \multicolumn{3}{c}{Public} &  \multicolumn{8}{c}{Routine}  &  \multicolumn{4}{c}{} \\
\cmidrule(r{0.7em}){3-5} \cmidrule(r{0.7em}l{0.7em}){6-13}  \multicolumn{15}{c}{} \\[-0.5em]
\toprule
Architecture & Trainingset & LTRC &  LCTSC &  VESS12 &  RRT &   Ath. &  Emph. &  Fibr. &  Mass &  PnTh &  Trau &  Norm &     All(L)* &        All &    All &   All \\
\midrule
\multirow{5}{*}{U-net} & R-36  & 0.99 &   0.93 &    0.98 &  0.92 &    0.95 &   0.99 &  0.96 &  0.98 &  0.99 &  0.93 &  0.97 &  \textbf{0.97$\pm$0.05} &  \textbf{0.96$\pm$0.08} &   \textbf{9.19$\pm$18.15} &  \textbf{1.43$\pm$2.26} \\
{} & LTRC-36  & 0.99 &   0.96 &    0.99 &  0.86 &   0.93 &   0.99 &   0.95 &  0.98 &  0.98 &  0.90 &  0.97 &  \textbf{0.97$\pm$0.08} &  0.94$\pm$0.13 &    11.9$\pm$22.9 &  2.42$\pm$5.99 \\
{} & LCTSC-36 &  0.98 &   0.97 &    0.98 & 0.85 &   0.91 &   0.98 &   0.92 &  0.98 &  0.98 &  0.89 &  0.97 &  0.96$\pm$0.09 &  0.92$\pm$0.14 &  10.96$\pm$14.85 &  1.96$\pm$2.87 \\
{} & VISC-36 & 0.98 &   0.95 &    0.98 &  0.84 &   0.91 &   0.98 &   0.90 &  0.98 &  0.98 &  0.89 &  0.97 &  0.96$\pm$0.09 &  0.92$\pm$0.15 &  13.04$\pm$19.04 &  2.05$\pm$3.08 \\ 
\midrule
\multicolumn{16}{c}{} \\[-2em]
\midrule
\multirow{5}{*}{ResU-net} & R-36  &   0.99 &   0.93 &    0.98 & 0.91 &  0.95 &   0.99 &   0.96 &  0.98 &  0.98 &  0.93 &  0.97 &  \textbf{0.97$\pm$0.06} &  \textbf{0.95$\pm$0.09} &   \textbf{8.66$\pm$15.06} &   \textbf{1.5$\pm$2.34} \\
{} & LTRC-36  &  0.99 &   0.96 &    0.99 & 0.86 &   0.94 &   0.99 &   0.95 &  0.98 &  0.98 &  0.89 &  0.97 &  \textbf{0.97$\pm$0.08} &  0.94$\pm$0.13 &  11.58$\pm$21.16 &  2.48$\pm$6.24 \\
{} & LCTSC-36 &  0.98 &   0.97 &    0.98 &  0.85 &  0.92 &   0.98 &   0.95 &  0.97 &  0.98 &  0.89 &  0.97 &  0.96$\pm$0.09 &  0.93$\pm$0.14 &  12.15$\pm$19.42 &  2.36$\pm$4.68 \\
{} & VISC-36 &   0.97 &   0.96 &    0.98 & 0.84 &  0.91 &   0.98 &   0.89 &  0.98 &  0.98 &  0.89 &  0.97 &  0.95$\pm$0.09 &  0.92$\pm$0.15 &    9.41$\pm$15.0 &  1.83$\pm$2.92 \\ 
\midrule
\multicolumn{16}{c}{} \\[-2em]
\midrule
\multirow{5}{*}{DRN} & R-36  &   0.98 &   0.93 &    0.97 & 0.88 &  0.94 &   0.98 &   0.95 &  0.97 &  0.98 &  0.92 &  0.96 &  \textbf{0.96$\pm$0.07} &  \textbf{0.94$\pm$0.12} &   8.96$\pm$17.67 &  1.96$\pm$3.97 \\
{} & LTRC-36  &  0.98 &   0.95 &    0.98 &  0.85 &  0.93 &   0.98 &   0.94 &  0.98 &  0.98 &  0.89 &  0.97 &  0.96$\pm$0.08 &  0.93$\pm$0.14 &  10.94$\pm$20.93 &  2.66$\pm$6.66 \\
{} & LCTSC-36 &   0.97 &   0.96 &    0.97 & 0.83 &  0.90 &   0.98 &   0.90 &  0.97 &  0.97 &  0.89 &  0.96 &  0.95$\pm$0.09 &  0.91$\pm$0.15 &    8.98$\pm$13.3 &  \textbf{1.92$\pm$2.73} \\
{} & VISC-36 &   0.96 &   0.95 &    0.97 & 0.83 &  0.90 &   0.97 &   0.92 &  0.97 &  0.97 &  0.87 &  0.97 &   0.94$\pm$0.1 &  0.91$\pm$0.15 &   \textbf{8.96$\pm$13.62} &  1.92$\pm$2.83 \\ 
\midrule
\multicolumn{16}{c}{} \\[-2em]
\midrule
\multirow{5}{*}{Deeplab v3+}& R-36  &   0.98 &   0.92 &    0.98 & 0.90 &  0.93 &   0.99 &   0.95 &  0.98 &  0.98 &  0.92 &  0.97 &  \textbf{0.96$\pm$0.06} & \textbf{ 0.95$\pm$0.09} &   \textbf{8.99$\pm$14.32} &  \textbf{1.71$\pm$2.68} \\
{} & LTRC-36  &   0.99 &   0.94 &    0.99 & 0.85 &  0.93 &   0.98 &   0.94 &  0.98 &  0.98 &  0.89 &  0.97 &  \textbf{0.96$\pm$0.09} &  0.93$\pm$0.14 &    11.9$\pm$21.8 &  2.51$\pm$6.07 \\
{} & LCTSC-36 &   0.98 &   0.96 &    0.98 & 0.85 &  0.92 &   0.98 &   0.93 &  0.98 &  0.98 &  0.89 &  0.96 &  \textbf{0.96$\pm$0.08} &  0.93$\pm$0.14 &  10.47$\pm$19.14 &  2.21$\pm$4.67 \\
{} & VISC-36 &   0.98 &   0.96 &    0.98 & 0.85 &  0.93 &   0.98 &   0.95 &  0.98 &  0.98 &  0.89 &  0.97 &  \textbf{0.96$\pm$0.08} &  0.93$\pm$0.14 &  10.16$\pm$21.21 &  2.15$\pm$4.99 \\ 
\bottomrule
\hspace{.05em}
\end{tabular}
\end{adjustbox}
\caption{\textbf{Evaluation results after training segmentation architectures on different training sets: }The sets R-36, LTRC-36, LCTSC-36, and LTRC-36 and VISC-36 contained the same number of volumes and slices. The best evaluation scores for models trained on these three datasets are marked in bold, highest for the Dice similarity score (DSC) and lowest for the Robust Hausdorff distance (HD95) and mean surface distance (MSD). Although the different architectures performed comparably, training on routine data outperformed training on public cohort datasets *The LCTSC ground truth masks do not include high-density areas, and the high number of LTRC test cases dominates the averaged results. Thus, “All(L)” (n = 167) is the mean over all cases including LCTSC and LTRC while “All” (n = 62) does not include the LCTSC or the LTRC cases. For abbreviations, see Tables 1 and 2}
\label{tbl:results_lonly}
\end{table*}

\begin{table*}[t]
\begin{adjustbox}{width=.99\textwidth,center}
\begin{tabular}{llrrrrrrrrrrllll}
{} & \multicolumn{11}{c}{Test datasets (DSC) for full volumes} &  \multicolumn{2}{c}{DSC$\pm$SD}  &    \multicolumn{1}{c}{HD95(mm)$\pm$SD} &   \multicolumn{1}{c}{MSD(mm)$\pm$SD} \\ \cmidrule(r{0.7em}){2-12} \cmidrule(r{0.7em}l{0.7em}){13-14} \cmidrule(r{0.7em}l{0.7em}){15-15} \cmidrule(r{0.7em}l{0.7em}){16-16}
{} &   \multicolumn{3}{c}{Public} &  \multicolumn{8}{c}{Routine}  &  \multicolumn{4}{c}{} \\
\cmidrule(r{0.7em}){2-4} \cmidrule(r{0.7em}l{0.7em}){5-12}  \multicolumn{15}{c}{} \\[-0.5em]
\toprule
Architecture &  LTRC &  LCTSC &  VESS12 & RRT & Ath. &  Emph. &  Fibr. &  Mass &  PnTh &  Trau &  Norm &     All(L)* &        All &   All(HD95) &   All(MSD) \\
\midrule
unet(R-231)    &  0.99 &   0.94 &    0.98 &  0.97 &  0.97 &   0.99 &      0.97 &  0.98 &  0.99 &  0.97 &  0.97 &  \textbf{0.98$\pm$0.03} &  \textbf{0.98$\pm$0.03} &   \textbf{ 3.14$\pm$7.4} &  \textbf{0.62$\pm$0.93} \\
resunet(R-231) &  0.99 &   0.94 &    0.98 &  0.97 &  0.97 &   0.99 &      0.97 &  0.98 &  0.99 &  0.97 &  0.97 & \textbf{0.98$\pm$0.03} &  \textbf{0.98$\pm$0.03} &   3.19$\pm$7.35 &  0.64$\pm$0.88 \\
drn(R-231)     &  0.98 &   0.94 &    0.98 &  0.95 &  0.96 &   0.99 &      0.97 &  0.98 &  0.98 &  0.96 &  0.97 &  0.97$\pm$0.04 &  0.97$\pm$0.06 &  6.22$\pm$18.95 &   1.1$\pm$2.54 \\
deeplab(R-231) &  0.99 &   0.94 &    0.98 &  0.97 &  0.97 &   0.99 &      0.97 &  0.98 &  0.99 &  0.97 &  0.97 &  \textbf{0.98$\pm$0.03} &  \textbf{0.98$\pm$0.03} &   3.28$\pm$7.52 &  0.65$\pm$0.91 \\
P-HNN         &  0.98 &   0.94 &    0.99 &  0.88 &  0.95 &   0.98 &      0.95 &  0.98 &  0.96 &  0.88 &  0.97 &  0.96$\pm$0.09 &  0.94$\pm$0.12 &  16.8$\pm$36.57 &  2.59$\pm$5.96 \\
\cmidrule{1-16}
unet(R-231)**   &  0.99 &   0.95 &    0.99 &  0.99 &  0.97 &   0.99 &      0.97 &  0.98 &  0.99 &  0.97 &  0.98 &  \textbf{0.98$\pm$0.01} &  \textbf{0.98$\pm$0.01} &   \textbf{1.44$\pm$1.09} &  \textbf{0.35$\pm$0.19} \\
CIP**          &  0.99 &   0.94 &    0.99 &  0.96 &  0.90 &   0.99 &      0.92 &  0.98 &  0.99 &  0.86 &  0.99 &  0.98$\pm$0.03 &  0.96$\pm$0.05 &   4.65$\pm$6.45 &  0.91$\pm$1.09 \\
CIP\#cases** & 96/105 & 19/24 & 17/20 & 13/20 & 2/2 & 2/2 & 4/4 & 2/2 & 2/2 & 1/3 & 1/7 & {} & {} & {} & {} \\
\bottomrule 
\hspace{.05em}
\end{tabular}
\end{adjustbox}
\caption{\textbf{Comparison to public systems:} A comparison to the segmentation algorithm of the chest imaging platform (CIP) and the trained P-HNN model is given. The results are expressed in mean and mean $\pm$ standard deviation for the Dice similarity coefficient (DSC), Robust Hausdorff distance (HD95), and mean surface distance (MSD) *The LCTSC ground truth masks do not include high-density diseases, and the high number of LTRC test cases dominates the averaged results. Thus, “All(L)” (n = 167) is the mean over all cases that included LCTSC and LTRC, while “All” (n = 62) does not include the LCTSC and LTRC cases **For these rows, only cases on which the CIP algorithm did not fail, and where the DSC was larger than 0 were considered (\#Cases). For abbreviations, see Tables 1 and 2}
\label{tbl:comp_models}
\end{table*}

\begin{table}[t]
\centering
\begin{tabular}{lrrrr}
\toprule
{} & \multicolumn{4}{c}{Tumor overlap} \\
Method &  Mean (\%) &  Median (\%)&  $<5\%$ &  $>95\%$ \\
\midrule
CIP         &  34 &    13 &    113 &     56 \\
P-HNN       &  50 &    44 &     48 &     78 \\
U-net(R-36) &  53 &    54 &     46 &     79 \\
U-net(R-231) &  \textbf{60} &    \textbf{69} &     \textbf{37} &    \textbf{ 90} \\
\bottomrule
\hspace{.05em}
\end{tabular}
\caption{Overlap between lung masks and manually annotated tumour volume in the Lung1 dataset. Given are mean, median, and number of cases with a smaller than 5\% overlap and a larger than 95\% overlap.}
\label{tbl:tumoroverlapp}
\end{table}

\section{Discussion}
We showed that training data, sampled from the clinical routine, improves generalizability to a wide spectrum of pathologies compared to public datasets. We assume this lies in the fact that many publicly available datasets do not include dense pathologies such as severe fibrosis, tumour, or effusions as part of the lung segmentation. Further, they are often provided without guarantees about segmentation quality and consistency. While the Anatomy3 dataset underwent a thorough quality assessment, the organisers of the VESSEL12 dataset merely provided lung segmentations as a courtesy supplement for the task of vessel segmentation, and within the LCTSC dataset, “tumour is excluded in most data” and “collapsed lung may be excluded in some scans” \cite{yang2018}. 
Results indicate that that both, size and diversity of the training data, are relevant. State-of-the-art results can be achieved with images from only 36 patients which is in line with previous works [41] achieving a mean DSC of 0.99 on LTRC test data using the U-net(R-36) model. 
A large number of segmentation methods are proposed every year, often based on architectural modifications \cite{isensee2019} of established models. Isensee et al. \cite{isensee2019} showed that such modified design concepts do not improve, and occasionally even worsen, the performance of a well-designed baseline. They achieved state-of-the-art performance on multiple, publicly available segmentation challenges relying only on U-nets. This corresponds to our finding that architectural choice had a subordinate effect on performance.
At the time of submission, the U-net(R-231) achieved the second highest score among all competitors in the LOLA11 challenge. In comparison, the first ranked method \cite{Soliman2017} achieved a score of 0.980 and a human reference segmentation achieved 0.984 \cite{vanRikxoortLObeLOLA11_}. Correspondingly, the U-net(R-231) model achieved improved evaluation measures (DSC, HD95, MSD and tumour overlap) compared to two public algorithms.
There are limitations of our study that should be taken into account. Routine clinical data vary between sites. Thus, extraction of a diverse training dataset from clinical routine may only be an option for centres that are exposed to a wide range of patient variety. Evaluation results based on public datasets are not fully comparable. For example, the models trained on routine data compared to other datasets yielded lower performance in terms of DSC on the LCTSC test data. However, the lower scores for models trained on routine data in LCTSC can be attributed to the lack of very-dense pathologies in the ground truth masks. Fig. \ref{fig:qual_pub} illustrates cases for which the R-231 model yielded low DSC. The inclusion or exclusion of pathologies such as effusions into lung segmentations is a matter of definition and application. While pleural effusions (and pneumothorax) are technically outside the lung, they are assessed as part of lung assessment, and have a substantial impact on lung parenchyma appearance through compression artefacts. Neglecting such abnormalities would hamper automated lung assessment, as they are closely linked to lung function. In addition, lung masks that include pleural effusions greatly alleviate the task of effusion detection and quantification, thus making it possible to remove effusions from the lung segmentation as a post-processing step. 
We proposed a general lung segmentation algorithm relevant for automated tasks in which the diagnosis is not known beforehand. However, specialized algorithms for specific diseases could be beneficial in scenarios of analyzing cohorts, for which the disease is already known.
In conclusion, we showed that accurate lung segmentation does not require complex methodology and that a proven deep-learning-based segmentation architecture yields state-of-the-art results once diverse (but not necessarily larger) training data are available. By comparing various datasets for training of the models, we illustrated the importance of training data diversity and showed that data from clinical routine can generalize well to unseen cohorts, highlighting the need for public datasets specifically curated for the task of lung segmentation. We draw the following conclusions: (1) translating ML approaches from bench to bedside can require the collection of diverse training data rather than methodological modifications; (2) current, publicly available study datasets do not meet these diversity requirements; and (3) generic, semantic, segmentation algorithms are adequate for the task of lung segmentation. A reliable, universal tool for lung segmentation is fundamentally important to foster research on severe lung diseases and to study routine clinical datasets. Thus, the trained model and inference code are made publicly available under the GPL-3.0 license to serve as an open science tool for research and development and as a publicly available baseline for lung segmentation under \url{https://github.com/JoHof/lungmask}.

\section*{Acknowledgment}
This research was supported by the Austrian Science Fund FWF (I2714-B31), Siemens Healthineers Digital Health (https://www.siemens-healthineers.com/digital-health-solutions), and an unrestricted grant of Boehringer Ingelheim.

\ifCLASSOPTIONcaptionsoff
  \newpage
\fi



\bibliographystyle{IEEEtran}
\bibliography{references2}{}

\begin{thebibliography}{10}
\providecommand{\url}[1]{#1}
\csname url@samestyle\endcsname
\providecommand{\newblock}{\relax}
\providecommand{\bibinfo}[2]{#2}
\providecommand{\BIBentrySTDinterwordspacing}{\spaceskip=0pt\relax}
\providecommand{\BIBentryALTinterwordstretchfactor}{4}
\providecommand{\BIBentryALTinterwordspacing}{\spaceskip=\fontdimen2\font plus
\BIBentryALTinterwordstretchfactor\fontdimen3\font minus
  \fontdimen4\font\relax}
\providecommand{\BIBforeignlanguage}[2]{{%
\expandafter\ifx\csname l@#1\endcsname\relax
\typeout{** WARNING: IEEEtran.bst: No hyphenation pattern has been}%
\typeout{** loaded for the language `#1'. Using the pattern for}%
\typeout{** the default language instead.}%
\else
\language=\csname l@#1\endcsname
\fi
#2}}
\providecommand{\BIBdecl}{\relax}
\BIBdecl

\bibitem{OECD2017}
{OECD}, ``{Health at a Glance 2017: OECD Indicators},'' Paris, 2017.

\bibitem{mansoor2015}
A.~Mansoor, U.~Bagci, B.~Foster, Z.~Xu, G.~Z. Papadakis, L.~R. Folio, J.~K.
  Udupa, and D.~J. Mollura, ``{Segmentation and Image Analysis of Abnormal
  Lungs at CT: Current Approaches, Challenges, and Future Trends},''
  \emph{RadioGraphics}, vol.~35, no.~4, pp. 1056--1076, 7 2015.

\bibitem{Zech2018}
J.~R. Zech, M.~A. Badgeley, M.~Liu, A.~B. Costa, J.~J. Titano, and E.~K.
  Oermann, ``{Variable generalization performance of a deep learning model to
  detect pneumonia in chest radiographs: A cross-sectional study},'' \emph{PLOS
  Medicine}, vol.~15, no.~11, p. e1002683, 11 2018.

\bibitem{goksel2015overview}
O.~G{\"{o}}ksel, O.~A. Jim{\'{e}}nez-del Toro, A.~Foncubierta-Rodr{\'{i}}guez,
  and H.~Muller, ``{Overview of the VISCERAL Challenge at ISBI},'' in
  \emph{Proceedings of the VISCERAL Challenge at ISBI}, New York, NY, 5 2015.

\bibitem{yang2018}
J.~Yang, H.~Veeraraghavan, S.~G. Armato, K.~Farahani, J.~S. Kirby,
  J.~Kalpathy-Kramer, W.~van Elmpt, A.~Dekker, X.~Han, X.~Feng, P.~Aljabar,
  B.~Oliveira, B.~van~der Heyden, L.~Zamdborg, D.~Lam, M.~Gooding, and G.~C.
  Sharp, ``{Autosegmentation for thoracic radiation treatment planning: A grand
  challenge at AAPM 2017},'' \emph{Medical Physics}, vol.~45, no.~10, pp.
  4568--4581, 10 2018.

\bibitem{Oakden2017}
L.~Oakden-Rayner, G.~Carneiro, T.~Bessen, J.~C. Nascimento, A.~P. Bradley, and
  L.~J. Palmer, ``{Precision Radiology: Predicting longevity using feature
  engineering and deep learning methods in a radiomics framework},''
  \emph{Scientific Reports}, vol.~7, no.~1, p. 1648, 2017.

\bibitem{Stein2016}
J.~M. Stein, L.~L. Walkup, A.~S. Brody, R.~J. Fleck, and J.~C. Woods,
  ``{Quantitative CT characterization of pediatric lung development using
  routine clinical imaging},'' \emph{Pediatric Radiology}, vol.~46, no.~13, pp.
  1804--1812, 12 2016.

\bibitem{Korfiatis2007}
P.~Korfiatis, S.~Skiadopoulos, P.~Sakellaropoulos, C.~Kalogeropoulou, and
  L.~Costaridou, ``{Combining 2D wavelet edge highlighting and 3D thresholding
  for lung segmentation in thin-slice CT},'' \emph{The British Journal of
  Radiology}, vol.~80, no. 960, pp. 996--1004, 12 2007.

\bibitem{Hu2001}
S.~Hu, E.~Hoffman, and J.~Reinhardt, ``{Automatic lung segmentation for
  accurate quantitation of volumetric X-ray CT images},'' \emph{IEEE
  Transactions on Medical Imaging}, vol.~20, no.~6, pp. 490--498, 6 2001.

\bibitem{Chen2011}
H.~Chen and A.~Butler, ``{Automatic Lung Segmentation in HRCT Images},'' in
  \emph{International Conference on Image and Vision Computing}, 2011, pp.
  293--298.

\bibitem{Pulagam2016}
A.~R. Pulagam, G.~B. Kande, V.~K.~R. Ede, and R.~B. Inampudi, ``{Automated Lung
  Segmentation from HRCT Scans with Diffuse Parenchymal Lung Diseases},''
  \emph{Journal of Digital Imaging}, vol.~29, no.~4, pp. 507--519, 8 2016.

\bibitem{Sluimer2005}
I.~Sluimer, M.~Prokop, and B.~van Ginneken, ``{Toward automated segmentation of
  the pathological lung in CT},'' \emph{IEEE Transactions on Medical Imaging},
  vol.~24, no.~8, pp. 1025--1038, 8 2005.

\bibitem{Iglesias2015}
J.~E. Iglesias and M.~R. Sabuncu, ``{Multi-atlas segmentation of biomedical
  images: A survey.}'' \emph{Medical image analysis}, vol.~24, no.~1, pp.
  205--19, 8 2015.

\bibitem{li2006}
Z.~Li, E.~A. Hoffman, and J.~M. Reinhardt, ``{Atlas-driven lung lobe
  segmentation in volumetric X-ray CT images},'' \emph{IEEE Transactions on
  Medical Imaging}, vol.~25, no.~1, pp. 1--16, 2005.

\bibitem{ShanhuiSun2012}
{Shanhui Sun}, C.~Bauer, and R.~Beichel, ``{Automated 3-D Segmentation of Lungs
  With Lung Cancer in CT Data Using a Novel Robust Active Shape Model
  Approach},'' \emph{IEEE Transactions on Medical Imaging}, vol.~31, no.~2, pp.
  449--460, 2 2012.

\bibitem{agarwala2017}
S.~Agarwala, D.~Nandi, A.~Kumar, A.~K. Dhara, S.~B. T.~A. Sadhu, and A.~K.
  Bhadra, ``{Automated segmentation of lung field in HRCT images using active
  shape model},'' in \emph{IEEE Region 10 Annual International Conference,
  Proceedings/TENCON}, vol. 2017-Decem.\hskip 1em plus 0.5em minus 0.4em\relax
  IEEE, 11 2017, pp. 2516--2520.

\bibitem{chen2019}
G.~Chen, D.~Xiang, B.~Zhang, H.~Tian, X.~Yang, F.~Shi, W.~Zhu, B.~Tian, and
  X.~Chen, ``{Automatic Pathological Lung Segmentation in Low-Dose CT Image
  Using Eigenspace Sparse Shape Composition},'' \emph{IEEE Transactions on
  Medical Imaging}, vol.~38, no.~7, pp. 1736--1749, 7 2019.

\bibitem{softka2011multistage}
M.~Sofka, J.~Wetzl, N.~Birkbeck, J.~Zhang, T.~Kohlberger, J.~Kaftan,
  J.~Declerck, and S.~K. Zhou, ``{Multi-stage learning for robust lung
  segmentation in challenging CT volumes},'' in \emph{International Conference
  on Medical Image Computing and Computer-Assisted Intervention}, vol. 6893
  LNCS, no. PART 3.\hskip 1em plus 0.5em minus 0.4em\relax Springer, Berlin,
  Heidelberg, 2011, pp. 667--674.

\bibitem{harrison2017}
A.~P. Harrison, Z.~Xu, K.~George, L.~Lu, R.~M. Summers, and D.~J. Mollura,
  ``{Progressive and multi-path holistically nested neural networks for
  pathological lung segmentation from CT images},'' in \emph{International
  Conference on Medical Image Computing and Computer-Assisted Intervention},
  vol. 10435 LNCS.\hskip 1em plus 0.5em minus 0.4em\relax Springer, Cham, 9
  2017, pp. 621--629.

\bibitem{Korfiatis2008}
P.~Korfiatis, C.~Kalogeropoulou, A.~Karahaliou, A.~Kazantzi, S.~Skiadopoulos,
  and L.~Costaridou, ``{Texture classification-based segmentation of lung
  affected by interstitial pneumonia in high-resolution CT},'' \emph{Medical
  Physics}, vol.~35, no.~12, pp. 5290--5302, 11 2008.

\bibitem{Wang2009}
J.~Wang, F.~Li, and Q.~Li, ``{Automated segmentation of lungs with severe
  interstitial lung disease in CT},'' \emph{Medical Physics}, vol.~36, no.~10,
  pp. 4592--4599, 9 2009.

\bibitem{Soliman2017}
A.~Soliman, F.~Khalifa, A.~Elnakib, M.~Abou El-Ghar, N.~Dunlap, B.~Wang,
  G.~Gimel'farb, R.~Keynton, and A.~El-Baz, ``{Accurate Lungs Segmentation on
  CT Chest Images by Adaptive Appearance-Guided Shape Modeling},'' \emph{IEEE
  Transactions on Medical Imaging}, vol.~36, no.~1, pp. 263--276, 1 2017.

\bibitem{Mansoor2014}
A.~Mansoor, U.~Bagci, Z.~Xu, B.~Foster, K.~N. Olivier, J.~M. Elinoff, A.~F.
  Suffredini, J.~K. Udupa, and D.~J. Mollura, ``{A Generic Approach to
  Pathological Lung Segmentation},'' \emph{IEEE Transactions on Medical
  Imaging}, vol.~33, no.~12, p. 2293, 12 2014.

\bibitem{Zhang2001}
\BIBentryALTinterwordspacing
Y.~Zhang, M.~Brady, and S.~Smith, ``{Segmentation of brain MR images through a
  hidden Markov random field model and the expectation-maximization
  algorithm},'' \emph{IEEE Transactions on Medical Imaging}, vol.~20, no.~1,
  pp. 45--57, 2001. [Online]. Available:
  \url{http://ieeexplore.ieee.org/document/906424/}
\BIBentrySTDinterwordspacing

\bibitem{rudyanto2012_}
R.~D. Rudyanto, S.~Kerkstra, E.~M. van Rikxoort, C.~Fetita, P.-Y. Brillet,
  C.~Lefevre, W.~Xue, X.~Zhu, J.~Liang, I.~Öksüz, D.~Ünay, K.~Kadipasaoglu,
  R.~S.~J. Estepar, J.~C. Ross, G.~R. Washko, J.-C. Prieto, M.~H. Hoyos,
  M.~Orkisz, H.~Meine, M.~Hüllebrand, C.~Stöcker, F.~L. Mir, V.~Naranjo,
  E.~Villanueva, M.~Staring, C.~Xiao, B.~C. Stoel, A.~Fabijanska, E.~Smistad,
  A.~C. Elster, F.~Lindseth, A.~H. Foruzan, R.~Kiros, K.~Popuri, D.~Cobzas,
  D.~Jimenez-Carretero, A.~Santos, M.~J. Ledesma-Carbayo, M.~Helmberger,
  M.~Urschler, M.~Pienn, D.~G. Bosboom, A.~Campo, M.~Prokop, P.~A. de~Jong,
  C.~Ortiz-de Solorzano, A.~Muñoz-Barrutia, and B.~van Ginneken, ``{Comparing
  algorithms for automated vessel segmentation in computed tomography scans of
  the lung: the VESSEL12 study},'' \emph{Medical Image Analysis}, vol.~18,
  no.~7, pp. 1217--1232, 10 2014.

\bibitem{vanRikxoortLObeLOLA11}
\BIBentryALTinterwordspacing
E.~M. van Rikxoort, B.~van Ginneken, and S.~Kerkstra, ``{LObe and Lung Analysis
  2011 (LOLA11)}.'' [Online]. Available:
  \url{ttps://lola11.grand-challenge.org}
\BIBentrySTDinterwordspacing

\bibitem{hofmanninger2016}
J.~Hofmanninger, M.~Krenn, M.~Holzer, T.~Schlegl, H.~Prosch, and G.~Langs,
  ``{Unsupervised identification of clinically relevant clusters in routine
  imaging data},'' in \emph{International Conference on Medical Image Computing
  and Computer-Assisted Intervention}, vol. 9900 LNCS, 2016, pp. 192--200.

\bibitem{itksnap}
P.~A. Yushkevich, J.~Piven, H.~C. Hazlett, R.~G. Smith, S.~Ho, J.~C. Gee, and
  G.~Gerig, ``{User-guided 3D active contour segmentation of anatomical
  structures: Significantly improved efficiency and reliability},''
  \emph{NeuroImage}, vol.~31, no.~3, pp. 1116--1128, 7 2006.

\bibitem{Ronneberger2015}
O.~Ronneberger, P.~Fischer, and T.~Brox, ``{U-net: Convolutional networks for
  biomedical image segmentation},'' in \emph{International Conference on
  Medical image computing and computer-assisted intervention}, vol. 9351, 2015,
  pp. 234--241.

\bibitem{Zhou2017}
\BIBentryALTinterwordspacing
X.~Zhou, R.~Takayama, S.~Wang, T.~Hara, and H.~Fujita, ``{Deep learning of the
  sectional appearances of 3D CT images for anatomical structure segmentation
  based on an FCN voting method},'' \emph{Medical Physics}, vol.~44, no.~10,
  pp. 5221--5233, 10 2017. [Online]. Available:
  \url{http://doi.wiley.com/10.1002/mp.12480}
\BIBentrySTDinterwordspacing

\bibitem{isensee2019}
F.~Isensee, J.~Petersen, S.~A.~A. Kohl, P.~F. J{\"{a}}ger, and K.~H.
  Maier-Hein, ``{nnU-Net: Breaking the Spell on Successful Medical Image
  Segmentation},'' \emph{arXiv preprint arXiv:1809.10486}, 4 2019.

\bibitem{batchnorm2015}
S.~Ioffe and C.~Szegedy, ``{Batch Normalization: Accelerating Deep Network
  Training by Reducing Internal Covariate Shift},'' in \emph{International
  Conference on Machine Learning}, 6 2015, pp. 448--456.

\bibitem{schmid2015}
R.~K. Srivastava, K.~Greff, and J.~Schmidhuber, ``{Training Very Deep
  Networks},'' in \emph{Advances in neural information processing systems},
  2015, pp. 2377--2385.

\bibitem{He2016}
K.~He, X.~Zhang, S.~Ren, and J.~Sun, ``{Deep Residual Learning for Image
  Recognition},'' in \emph{Proceedings of the IEEE conference on computer
  vision and pattern recognition}, 2016, pp. 770--778.

\bibitem{yu2015}
F.~Yu and V.~Koltun, ``{Multi-Scale Context Aggregation by Dilated
  Convolutions},'' 11 2015.

\bibitem{yu2017}
F.~Yu, V.~Koltun, and T.~Funkhouser, ``{Dilated Residual Networks},''
  \emph{Proceedings of the IEEE conference on computer vision and pattern
  recognition}, pp. 472--480, 2017.

\bibitem{Chen2018}
L.-C. Chen, G.~Papandreou, I.~Kokkinos, K.~Murphy, and A.~L. Yuille,
  ``{DeepLab: Semantic Image Segmentation with Deep Convolutional Nets, Atrous
  Convolution, and Fully Connected CRFs},'' \emph{IEEE Transactions on Pattern
  Analysis and Machine Intelligence}, vol.~40, no.~4, pp. 834--848, 4 2018.

\bibitem{ChestCIP_}
\BIBentryALTinterwordspacing
{Chest Imaging Platform (CIP)}. [Online]. Available:
  \url{https://chestimagingplatform.org}
\BIBentrySTDinterwordspacing

\bibitem{Aerts2014}
H.~J. Aerts, E.~R. Velazquez, R.~T. Leijenaar, C.~Parmar, P.~Grossmann,
  S.~Carvalho, J.~Bussink, R.~Monshouwer, B.~Haibe-Kains, D.~Rietveld,
  F.~Hoebers, M.~M. Rietbergen, C.~R. Leemans, A.~Dekker, J.~Quackenbush, R.~J.
  Gillies, and P.~Lambin, ``{Decoding tumour phenotype by noninvasive imaging
  using a quantitative radiomics approach},'' \emph{Nat Commun}, vol.~5, p.
  4006, 2014.

\bibitem{2018LibraryTasks}
\BIBentryALTinterwordspacing
``{Library to compute surface distance based performance metrics for
  segmentation tasks},'' 2018. [Online]. Available:
  \url{https://github.com/deepmind/surface-distance}
\BIBentrySTDinterwordspacing

\bibitem{vanRikxoortLObeLOLA11_}
\BIBentryALTinterwordspacing
E.~M. van Rikxoort, B.~van Ginneken, and S.~Kerkstra. {LObe and Lung Analysis
  2011 (LOLA11)}. [Online]. Available: \url{https://lola11.grand-challenge.org}
\BIBentrySTDinterwordspacing

\end{thebibliography}

\end{document}